\documentclass[useAMS,usenatbib]{mn2e}
\usepackage{graphicx}
\usepackage{epsfig}

\title[{\it Suzaku} observation of G304.6$+$0.1]{A deep X-ray observation of supernova remnant G304.6$+$0.1 (Kes 17) with {\it Suzaku}}
\author[F.~G\"{o}k, A. Sezer]{F.~G\"{o}k,$^{1}$\thanks{E-mail: gok@akdeniz.edu.tr
(FG); aytap.sezer@boun.edu.tr (AS)} A. Sezer $^{2,3}$
\footnotemark[1]\thanks{This file has been amended to highlight
the proper use of \LaTeXe\ code with the class file. These changes
are for illustrative purposes and do not reflect the
original paper by F.~G\"{o}k.}\\
$^{1}$Akdeniz University, Faculty of Education, Antalya, 07058, Turkey\\
$^{2}$T\"UB\.ITAK Space Technologies Research Institute, ODTU
Campus, Ankara, 06531, Turkey\\
$^{3}$Bo\~gazi\d{c}i University, Faculty of Art and Sciences,
Department of Physics, \.Istanbul, 34342, Turkey\\
}
\begin{document}

\date{}

\pagerange{\pageref{firstpage}--\pageref{lastpage}} \pubyear{2011}

\maketitle

\label{firstpage}

\begin{abstract}
In this paper, we present the analysis of a deep (99.6 ks)
observation of G304.6$+$0.1 with the X-ray Imaging Spectrometer on
board {\it Suzaku} satellite. The X-ray spectral data are
well-fitted with a plasma model consisting of a thermal component
in collisional ionization equilibrium and a non-thermal component.
The thermal emission is well fitted with VMEKAL model with an
electron temperature of $kT_{\rm e}\sim 0.75$ keV, a high
absorbing column density of $N_{\rm H}\sim 3.9\times10^{22}$ $\rm
cm^{-2}$ and near/lower solar abundances which indicate that the
X-ray emitting plasma of G304.6$+$0.1 is dominated by swept-up
ambient medium. The non-thermal component is well fitted with a
power-law model with photon index of $\Gamma \sim 1.4$. We found a
relatively high electron density $n_{\rm e}\sim 2.3f^{-1/2}$
cm$^{-3}$, age $t$ $\sim 1.4\times10^4f^{1/2}$ yr, and X-ray
emitting mass $M_{\rm x}\sim 380f^{1/2}$ {M\sun} at an adopted
distance of d=10 kpc. Using the morphological and spectral X-ray
data, we confirm that the remnant is a new member of
mixed-morphology supernova remnants.

\end{abstract}


\section{Introduction}

Recent supernova remnant (SNR) observations in X-ray band led to
discovery of a considerable number of a new morphological class
called mixed-morphology (MM) SNRs that have a centrally bright in
the X-ray band characterized by a thermal spectrum with little or
no limb brightening and a shell morphology in radio \citep {b7}.
Their characteristics have been explained with two basic
scenarios; evaporating clouds \citep {b26, b9} and thermal
conduction models \citep {b27, b28, b29}. In addition to these
basic ones there are new scenarios that try to explain the
mechanism taking place, but MM SNRs are still not understood
completely. Deep X-ray observations will help understanding the
mechanism taking place in these MM SNRs.

G304.6+0.1 (Kes 17), a new MM member \citep {b4}, was first
detected at 408 MHz and 5 GHz with a Parkes and a Molonglo radio
telescopes by \citet {b14}. \citet {b15} gave a lower limit of 9.7
kpc to the distance of the remnant by measuring 21 cm absorption
profiles with the Parkes hydrogen line interferometer. From the
radio observations, \citet {b35} suggested that its structure has
been resulted from the shock front collision with its dense
environment, \citet {b6} have shown that it had a irregular shell
morphology with a small size. \citet {b16} have detected 1720 MHz
OH maser emission toward G304.6+0.1 by using a Parkes Telescope.
\citet {b18}, using mid-infrared data from the Spitzer Space
Telescope at 3.6, 4.5, 5.8 and 8 $\mu$m showed that the remnant
has a shell structure in infrared and detected a pair of thin
filament at about $\rmn{RA}(2000)=13^{\rmn{h}} 05^{\rmn{m}}
46^{\rmn{s}}.2$, $\rmn{Dec.}(2000)=-62\degr 38\arcmin 33\arcsec$.
The authors claimed that most of the infrared emission coming from
the shell could be resulted from shocked molecular material.
\citet {b30} have obtained sensitive low-resolution spectroscopy
from 5 to 35 $\mu$m for this remnant with Spitzer IRS and showed
evidence of shocked molecular gas. The first detailed
multi-wavelength (radio, mid-infrared, and X-ray) study of the
G304.6$+$0.1 has been conducted by \citet {b4}. By using {\it
XMM-Newton} data (26.8 ks) they have found that the X-ray emission
was dominated by thermal radiation (PSHOCK) plus a small fraction
of non-thermal radiation (power-law) in north, center and south
regions. They have found a high column density $N_{\rm H}$ in the
range of ($2.5-3.5$)$\times10^{22}$ $\rm cm^{-2}$, $kT_{\rm e}$
$\sim 0.75$ keV and plasma was in non-equilibrium ionization state
with $\tau\sim2.1\times10^{12}$ $\rm cm^{-3}$s. In addition, they
calculated electron density $n_{\rm e}$ to be $\sim$0.99,
$\sim$0.89 and $\sim$2.26 cm$^{-3}$ for north, center and south
regions, respectively. Their multi-wavelength study has suggested
that this remnant was a middle-aged SNR, and belonged to the MM
SNRs class. \citet {b22} showed that the IR emission was
concentrated on the western and southern shells by using data
obtained with $\it AKARI$ and $\it Spitzer$ satellites, the
western shell was bright in the mid and far-IR continuum and the
near-IR H$_{2}$ line emission, whereas the southern shell was
visible only in the mid-IR continuum emission. The authors
suggested that far-IR continuum and near-IR H$_{2}$ line emission
emanating from the western shell was resulted from the interaction
with nearby molecular cloud. \citet {b17} reported the discovery
of GeV emission at the position of this remnant by using the data
from the Large Area Telescope on board the {\it Fermi} Gamma-ray
Space Telescope. They observed the remnant in the 1$-$20 GeV range
with a significance of $\sim$12$\sigma$, and suggested that Kes 17
was a candidate acceleration site for cosmic-rays, and reported
the detection of a number of ${\gamma}$-ray sources in its
vicinity.

{\it Suzaku} \citep {b12}, the newest Japanese X-ray astronomy
satellite that has a good energy resolution, high efficiency, and
low background level for diffuse sources, has observed
G304.6$+$0.1 with 99.6 ks exposure time. Our aim is to study the
remnant whose morphology implies a similarity to the recently
proposed group of MM SNRs by using the high quality spectral data.

The rest of this paper is structured as follows. Observation and
data reduction are described in Section 2, image and spectral
analysis are reported in Section 3. We discuss results of the
spectral analysis and finally, investigate the X-ray
characteristics of G304.6$+$0.1 from radial spectral variations
and compared the results with the two basic scenarios explaining
MM SNRs in Section 4.

\section[]{Observation and Data Reduction}
{\it Suzaku} observed G304.6$+$0.1 with the X-ray imaging
spectrometer (XIS: \citet {b19}) on 2010 September 03, for 99.6 ks
(Obs ID:505074010). {\it Suzaku} has three working XIS CCDs, two
of the cameras (XIS0 and XIS3) have front-illuminated (FI) CCDs,
and the remaining one (XIS1) has a back-illuminated (BI) CCD. Each
CCD camera has a single CCD chip with an array of $1024\times1024$
pixels, and covers an area of 17.8$\times$17.8 arcmin$^{2}$. The
XIS was operated with normal clocking mode, in $5\times5$ and
$3\times3$ editing modes.

We used the {\sc headas} 6.5 sofware package for data reduction.
The XIS data have been screened with {\sc xselect} using the
standard criterion \citep {b19}. Spectral analysis and model
fitting have been performed with {\sc xspec} v11.3 \citep {b2}.
The redistribution matrix files (RMFs) of the XIS has been
produced by xisrmfgen and auxillary responce files (ARFs) by
xissimarfgen \citep {b5}.

\section{Analysis}
\subsection{Image analysis}
Fig. 1 shows XIS0 image of G304.6$+$0.1 in 0.3$-$10 keV full
energy band. The outer most ellipse represents the over all region
(we call region 1) that X-ray emission originates. Inside the
region 1, the emission looks much concentrated in two parts, inner
ellipse (we call region 2) and inner circle (we call region 3).
The dashed circle and dashed square show background region and the
XIS field of view (FOV), respectively. Fig. 2 shows the XIS0 image
in 0.3$-$10 keV band overlaid with contours of the 843 MHz radio
image from \citet {b6}.

\subsection{Spectral analysis}

We extracted the spectrum for each XIS detector for the outer
elliptical region, inner elliptical region and inner circular
region. The background has been extracted from circular region,
taking care not to include the $^{55}$Fe calibration sources in
the corners of the CCD. In Table 1, we gave details of ellipses,
circles and background (centre coordinates, sizes and angles). The
spectra are grouped with a minimum of 50 counts bin$^{-1}$ for
region 1, 30 counts bin$^{-1}$ for region 2 and 25 counts
bin$^{-1}$ for region 3, and the $\chi^{2}$ statistics is used.

For spectral fitting of region 1, we first applied single
temperature non-equilibrium ionization (NEI) model (VNEI in {\sc
xspec}; \citet {b3}) modified by Galactic absorption via the WABS
multiplicative model \citep {b13}. The parameters of the absorbing
column density ($N_{\rm H}$), electron temperature ($kT_{\rm e}$)
and ionization parameter ($\tau=n_{\rm e}t$) were set free while
all elemental abundances were fixed at their solar values \citep
{b1}. The model gave a large ionization parameter of
$\tau\sim10^{13}$ $\rm cm^{-3}$s, which means that the plasma of
the remnant has reached ionization equilibrium with a reduced
$\chi^{2}$ of 1.43 (948/664 d.o.f.). Therefore, we tried an
absorbed collisional ionization equilibrium (CIE) plasma model
(VMEKAL in {\sc xspec}; \citet {b10, b11, b8}). In this case, the
reduced $\chi^{2}$ value (926/665=1.39) improved a little bit but
it was clear that the model fitting needed a second component.
Therefore, we added a second thermal component to VMEKAL, a CIE
model (VMEKAL) and a NEI model (VNEI), respectively. From the
model fitting of VMEKAL+VMEKAL we did not see two different
temperatures, but when we thawed Mg, Si, S we obtained two
different temperatures of $\sim$0.6 and $\sim$1.1 keV with a
reasonably good reduced $\chi^{2}$ of 1.15 (761.7/660 d.o.f.).
However, we obtained unphysically low abundance of S and very high
errors of abundances of Mg, Si and S. From the model fitting of
VMEKAL+VNEI, we obtained two different temperatures of $\sim$0.6
and $\sim$1.4 keV. When we thawed Mg, Si, S, the $\chi^{2}$ value
improved significantly (750.8/659=1.14). For both model fittings
we used F-test to test whether the extra thermal component was a
statistically significant addition. However, the F-test gave a
probability of $\sim$$1.5\times10^{-7}$ and
$\sim$$6.9\times10^{-9}$ that is the improvement is due to chance
for VMEKAL+VMEKAL and VMEKAL+VNEI models, respectively. Then, we
added a power-law component to VMEKAL model. In this case, we
obtained a reasonably good physical parameter values and their
errors with an acceptable $\chi^{2}$ value (799/663=1.2). Then, we
let Mg, Si and S vary which yielded a rather good reduced
$\chi^{2}$ value of 1.06 (700/660 d.o.f.). We also applied F-test
to our model fitting which gives a probability of
$\sim$$1.2\times10^{-19}$ that the improvement is not due to
chance. All these steps were repeated for region 2 and region 3.
For region 2, VMEKAL model required a power-law component.
However, region 3 did not require a second component. The XIS0,
XIS1 and XIS3 spectra of three regions were fitted simultaneously
in 0.3$-$10 keV energy band and are given in Fig. 3 (a, b and c).
The right panels of this figure show the individual components of
the applied model, VMEKAL and power-law, for XIS1 only.
Best-fitting parameters of applied models for three regions are
listed in Table 2. All the errors are at 90 per cent confidence
level.

We also studied spectral variation across the remnant and gave the
results in Fig. 4. To study the radial variation of $kT_{\rm e}$
and surface brightness we chose five elliptical annuli with
$0.6\times1.4$, $1.3\times2.1$, $1.9\times2.8$, $2.7\times3.6$ and
$3.6\times4.7$ arcmin$^{2}$ regions centered at
$\rmn{RA}(2000)=13^{\rmn{h}} 05^{\rmn{m}} 48^{\rmn{s}}$,
$\rmn{Dec.}(2000)=-62\degr 41\arcmin 46\arcsec$ from XIS0 FOV. The
spectra extracted from the elliptical annuli were grouped with a
minimum of 20 counts bin$^{-1}$. We applied our best-fitting model
(an absorbed VMEKAL and power-law) starting from the central
region within $0.6\times1.4$ arcmin$^{2}$, we let absorbing column
density, electron temperature, normalization and abundances of Mg,
Si and S vary. For all outer annuli, we applied same model fixing
$N_{\rm H}$ to the value obtained from central region.

\section{Discussion and Conclusions}
In this work, we have studied X-ray morphology and spectral
properties of G304.6$+$0.1 using XIS data obtained with deep
observation of {\it Suzaku}. As seen from Fig. 2, G304.6$+$0.1 has
a centrally filled X-ray morphology dominated by thermal X-ray
emission, the radio contours of remnant are concentrated at its
shell. Morphologically, this remnant resembles MM SNRs. With the
help of high quality spectral data we studied spectral
characteristics of G304.6$+$0.1 and compared our findings with the
theoretical models proposed for MM SNRs below.

The X-ray emission from the remnant is well described by a model
of a thermal plasma that has reached ionization equilibrium and a
non-thermal component (power-law) required to describe the plasma
state has a photon index ($\Gamma$) of $\sim$1.4. Dominant
emission from this remnant is thermal ($\sim$90 per cent of the
total X-ray flux) and is described by the CIE plasma model (in CIE
plasma, $n_{\rm e}t$ should be larger than $10^{12}$ $\rm
cm^{-3}$s \citep {b20}) with an electron temperature of $kT_{\rm
e}\sim0.75$ keV. We have performed detailed spectral fitting for
three regions. As seen from Table 2, spectral parameters for three
regions are almost similar. We obtained a high absorbing column
density $N_{\rm H}$ in the range of ($3.2-4.2$)$\times10^{22}$
$\rm cm^{-2}$ which is consistent with the previous work of \citet
{b4}, in the range of ($2.5-3.5$)$\times10^{22}$ $\rm cm^{-2}$,
that supported the large distance d $\geq 9.7$ kpc adopted by
\citet {b15} to the remnant. Our $N_{\rm H}$ value is larger than
that of \citet {b4} therefore, throughout our calculations we used
d=10 kpc. From the emission measure ($EM=n_{\rm e}n_{\rm H}V$,
where $n_{\rm e}$ is the electron density, $n_{\rm H}$ is the
hydrogen density, and $V$ is the volume of the X-ray emitting
plasma) we estimate the electron density of the plasma taking into
account $n_{\rm e}=1.2n_{\rm H}$ for the mean charge state with
normal composition. Considering the possibility that less than the
entire volume is filled, we calculate the volume of the X-ray
emitting region from $V=\frac{4}{3}\pi R^{3}f$, where $R$ is the
SNR radius and $f$ is the filling factor. Then we carry the $f$
factor through our calculations to see the dependence of each
derived quantity (for example the electron density, age) on $f$.
We obtain the volume to be $\sim$$1.9\times10^{59}f$ ${\rm
cm^{3}}$ assuming the spherical radius of 4 arcmin at the distance
d=10 kpc and then the electron density to be $\sim$$2.3f^{-1/2}$
cm$^{-3}$. In calculating the age of the remnant we use the
relation $\tau/n_{\rm e}$, where $\tau$ is the ionization age of
the plasma. CIE model gives only a lower limit ($10^{12}$
cm$^{-3}$s) to this parameter and hence leads to a lower limit on
the age $t$ of the remnant which is calculated to be
$\sim$$1.4\times10^4f^{1/2}$ yr for G304.6$+$0.1. The mass of the
X-ray emitting gas is calculated to be $\sim$$380 f^{1/2}$ {M\sun}
from $M_{\rm x}=m_{\rm H}n_{\rm e}V$, where $m_{\rm H}$ is mass of
a hydrogen atom. As seen from Table 2, near/lower solar abundance
values obtained for each region and large X-ray emitting mass
indicate that the X-ray emission results from swept-up ambient
medium as is generally the case in middle-aged SNRs. Relatively
large $n_{\rm e}$ ($\sim$$2.3f^{-1/2}$ cm$^{-3}$) and small size
($\sim$8 arcmin) suggest that the remnant expands in a region with
a large ambient density.

For the non-thermal component that constituting $\sim$10 per cent
of the total flux, our best-fitting photon index is in the range
of 1.4$-$1.6, as seen in Table 2, which is consistent with that of
classical young pulsars ranging in between 1.1 and 1.7 \citep
{b21}. But there is no compact or extended X-ray source reported
at the location or in the vicinity of this remnant. Another
possibility would be that the power-law emission is resulted from
the synchrotron emission from shock-accelerated relativistic
electrons, also as the {\it Fermi} detection of G304.6$+$0.1
\citep {b17} appears to be related to the remnant and not to a
central compact object. However, our spectral fitting of each
elliptical annulus gives different photon index value ({$\Gamma$
$\sim 1.5$, 1.2, 1.3, 1.5}), moreover the photon index obtained
from the centre is unphysically large ({$\Gamma$ $\sim 7.3$})
indicating that the power-law emission can not be related to the
remnant itself. That our best-fitting model required no
non-thermal component for region 3 may support this idea.
Considering all these cases, it is still difficult to explain the
origin of the non-thermal emission with {\it Suzaku} data. \citet
{b4} also obtained non-thermal emission from this remnant with a
photon index value in the range of 1.8$-$3.1 with {\it XMM-Newton}
observation, which is softer than our photon index value,
indicating particle acceleration in shock-fronts.

In addition to morphological evidences, G304.6$+$0.1 is a
middle-aged ($\sim$$10^{4}$ yr) SNR, the plasma of the remnant is
in CIE condition and it is located in an inhomogeneous medium. The
general characteristics of MM SNRs have been given in \citet {b7}
as (i) Members of this class have a centrally peaked morphology in
the X-ray band characterized by a thermal spectrum while a shell
morphology in radio, (ii) Their ages are expected to be
$\sim$$10^{4}$ yr, which means that the plasma should be in CIE
condition, (iii) They are interacting with molecular clouds and/or
HI clouds, which means that they are located in a very dense
medium. Considering these properties we may say that G304.6$+$0.1
is likely to be a member of MM SNRs as suggested by \citet {b4}.
The two basic scenarios explaining this class are; one is that the
interior X-ray emission arises from the gas evaporated from
shocked clouds \citep {b26, b9}. The second one is that as an SNR
evolves, the temperature and density of the hot interior plasma
gradually become uniform through thermal conduction and the
temperature in the outer shells becomes lower and absorbed by the
interstellar medium, the only detectable X-ray emission emanates
from the interior of the SNR \citep {b27, b28, b29}. To compare
our results with these models, we have studied radial variations
of the $kT_{\rm e}$ and the surface brightness as seen from Fig.
4. We have found that the temperature variation ($kT_{\rm e}$
$\sim$ 0.66$-$0.81 keV) is almost uniform across the remnant as
predicted by both models. On the contrary, surface brightness
peaks at the center and declines towards shell regions from
$\sim$2.34$\times10^{-12}$ erg $\rm s^{-1}$$\rm cm^{-2}$$\rm
arcmin^{-2}$ to $\sim$0.03$\times10^{-12}$ erg $\rm s^{-1}$$\rm
cm^{-2}$$\rm arcmin^{-2}$ in 0.3$-$10 keV, which is consistent
with evaporating clouds model. G304.6$+$0.1 is interacting with an
OH maser source at 1720 MHz \citep {b16}. Recently \citet {b22}
suggested that the remnant was interacting with nearby molecular
cloud. Furthermore, \citet {b17} have detected GeV emission from
the SNR and a number of $\gamma$-ray sources in its vicinity. The
thermal conduction model requires a relatively high density
ambient medium, while the evaporation model requires dense clouds.
In this respect, G304.6$+$0.1 is consistent with both models.

We present here the results from the {\it Suzaku} archival data of
G304.6$+$0.1. The deep XIS spectra consist of two components;
thermal emission coming from the plasma is in CIE condition with
$kT_{\rm e}\sim 0.75$ keV, a high $N_{\rm H}$ value and the
non-thermal emission has a photon index of $\sim$1.4. This value
is consistent with that of classical young pulsars, but there is
no observed compact source reported. So, the origin of the
non-thermal emission is still unclear. Thermal emission originates
from the shocked ambient medium with an electron density of
$n_{\rm e}\sim 2.3f^{-1/2}$ cm$^{-3}$. The morphological and X-ray
spectral characteristics imply that G304.6$+$0.1 is a member of MM
class.

\section*{Acknowledgments}
We thank to the referee for his/her detailed and constructive
comments and suggestions on the manuscript. We also acknowledge
support by the Akdeniz University Scientific Research Project
Management. AS is supported by T\"{U}B\.{I}TAK PostDoctoral
Fellowship.

\begin{table*}
 \begin{minipage}{140mm}
  \caption{Centre coordinates, sizes and angles of region 1, region 2, region 3 and background.}
 \begin{tabular}{@{}cccc@{}}
  \hline
     Regions & Centre Coordinates
& Size  & Angle \\
& $\rmn{RA}(2000)$ , $\rmn{Dec.}~(2000)$
& (arcmin)  & (degrees) \\
& ($~^{\rmn{h}} ~^{\rmn{m}} ~^{\rmn{s}}$ , $~\degr ~\arcmin
~\arcsec$)
&  &  \\
 \hline
1 & 13 05 44 , $-$62 41 54 & $3.2\times 4.1$ & $354$ \\
2 & 13 05 55 , $-$62 41 11 & $1.9\times 3.4$ & $342$ \\
3 & 13 05 35 , $-$62 43 33 &  $1.24$ & $0$ \\
Background &13 04 52 , $-$62 37 16 & $1.56$ & $0$ \\
\hline
\end{tabular}
\end{minipage}
\end{table*}

\begin{table*}
\centering
 \begin{minipage}{140mm}
  \caption{Best-fitting parameters with corresponding errors at 90
   per cent confidence level in 0.3$-$10 keV
energy band.}
 \begin{tabular}{@{}ccccccc@{}}
  \hline
      Component & Parameters & Region 1& Region 2& Region 3& \\
 \hline
 Wabs& $N_{\rm H}$($\times10^{22}$$\rm cm^{-2})$ &3.9$\pm 0.1$&4.2$\pm 0.1$& 3.2$\pm 0.1$&  \\
VMEKAL& $kT_{\rm e}$(keV) & 0.75 $\pm 0.01$&0.74 $\pm 0.01$ &0.84 $\pm 0.03$ & \\
Abundance\footnote{Abundances are relative to the solar values
\citep{b1}.} & Mg & 1.3 $\pm 0.2$&1.2 $\pm 0.2$&0.9 $\pm 0.2$ &  \\
 & Si  & 0.9 $\pm 0.1$&0.9 $\pm 0.1$&1.1 $\pm 0.1$ & \\
 & S  & 0.6 $\pm 0.1$&0.5 $\pm 0.1$& 0.4 $\pm 0.1$& \\
& norm (photons $\rm cm^{-2}s^{-1}$) & 0.93 $\pm 0.06$&0.65 $\pm 0.11$ &0.07 $\pm 0.01$ & \\
&$EM$\footnote{Emission measure $EM=\int n_{\rm e}n_{\rm H}$dV in
the unit of $10^{59}$ $\rm cm^{-3}$, where $n_{\rm e}$ and $n_{\rm
H}$ are number densities of electrons and protons, respectively
and V is
the X-ray emitting volume.}&11.1 $\pm 0.8$&7.7 $\pm 1.1$&0.8 $\pm 0.1$& \\
&Flux\footnote{Absorption-corrected thermal flux in $0.3-10$ keV energy band in the unit of $10^{-11}$ erg $\rm s^{-1}$$\rm cm^{-2}$.} &4.6 $\pm 0.1$ &3.2 $\pm 0.2$&0.6 $\pm 0.1$&\\
Power-law & Photon Index &1.4 $\pm 0.1$ &1.6 $\pm 0.2$&$-$& \\
 & norm ($\times10^{-2}$photons $\rm cm^{-2}s^{-1}$) & 1.17 $\pm 0.23$& 0.46 $\pm 0.21$& $-$& \\
&Flux\footnote{Absorption-corrected total flux (VMEKAL plus power-law components) in $0.3-10$ keV energy band in the unit of $10^{-11}$ erg $\rm s^{-1}$$\rm cm^{-2}$.} &5.5 $\pm 0.2$ &3.6 $\pm 0.2$&$-$&\\
 & $\chi^{2}$/d.o.f.  &700/660=1.06 &747/657=1.14&253.8/192=1.32 & \\
\hline
\end{tabular}
\end{minipage}
\end{table*}

\onecolumn

\begin{figure}
\centering
  \vspace*{17pt}
  \includegraphics[width=6cm]{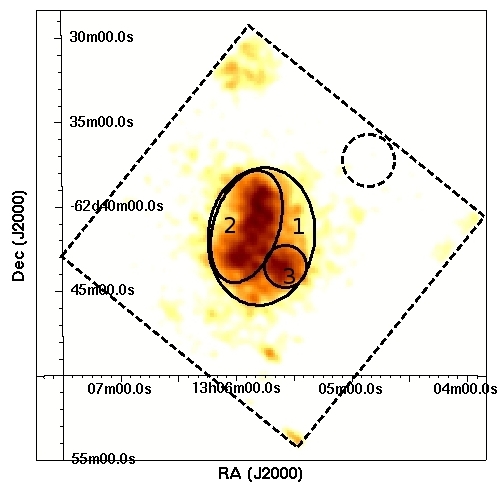}
  \caption{{\it Suzaku} XIS0 image of G304.6$+$0.1 in 0.3$-$10 keV energy
  band. The outer and inner ellipses represent region 1 and region 2 respectively and
inner circle represents region 3. The dashed circle and dashed
square show background region and FOV of XIS0, respectively.}
\end{figure}

\begin{figure}
\centering
  \vspace*{17pt}
\includegraphics[width=6cm]{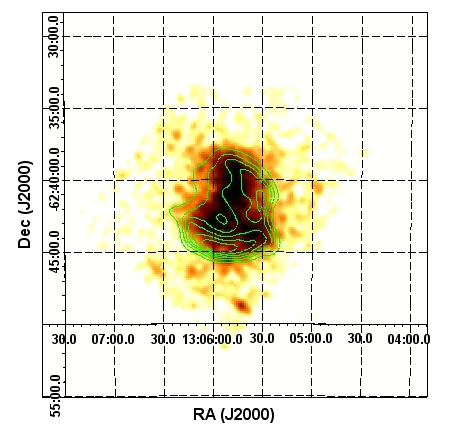}
\caption{{\it Suzaku} XIS0 image of G304.6$+$0.1 in 0.3$-$10 keV
full energy band overlaid with contours of the 843 MHz radio image
from \citet {b6}. The X-ray image is smoothed with a Gaussian
kernel of $\sigma$=2 arcsec, and units are in counts s$^{-1}$.
Overlaid contours are spaced linearly in intensity of 0.07, 0.26,
0.44 and 0.62 count pixel$^{-1}$.}
\end{figure}

\begin{figure}
\centering
  \vspace*{17pt}
\includegraphics[width=8cm]{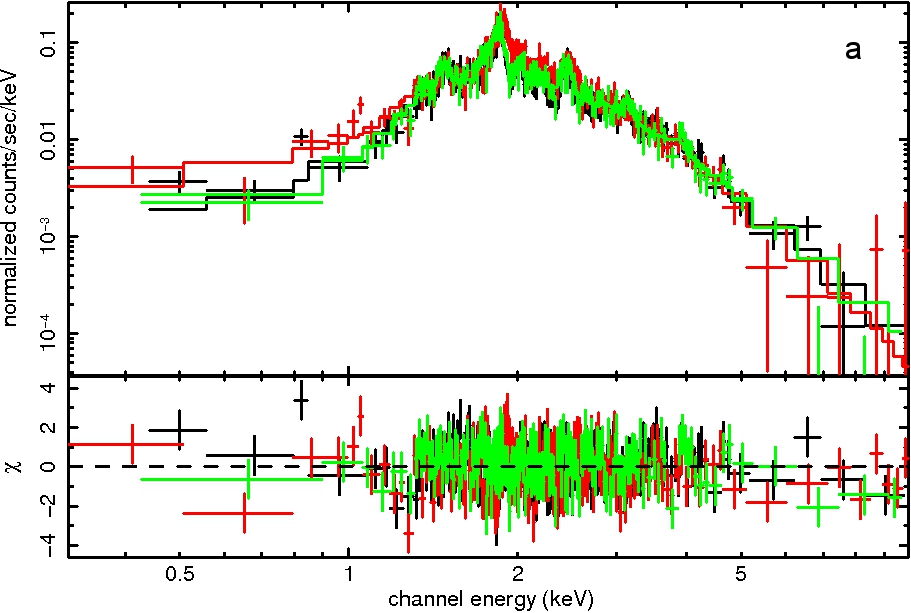}
\includegraphics[width=8cm]{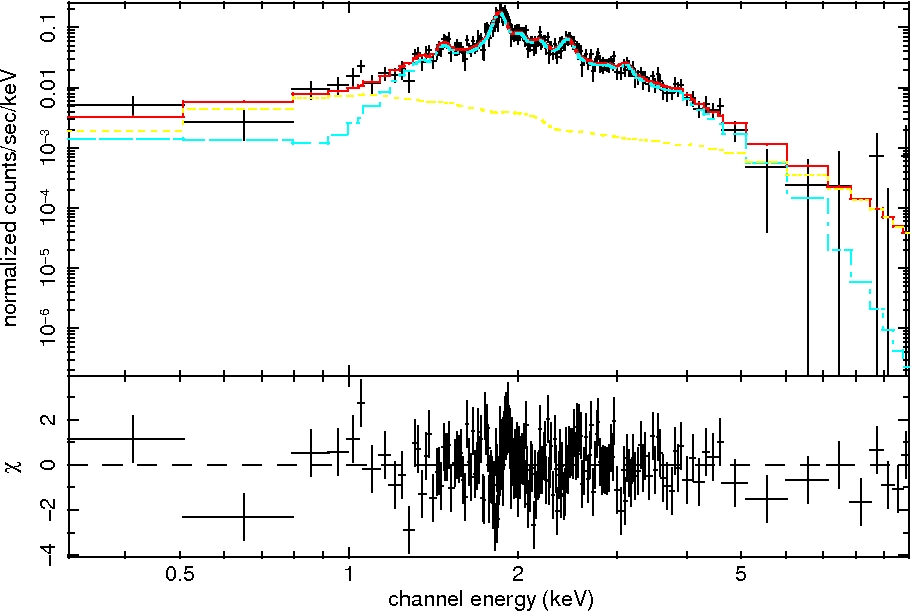}
\includegraphics[width=8cm]{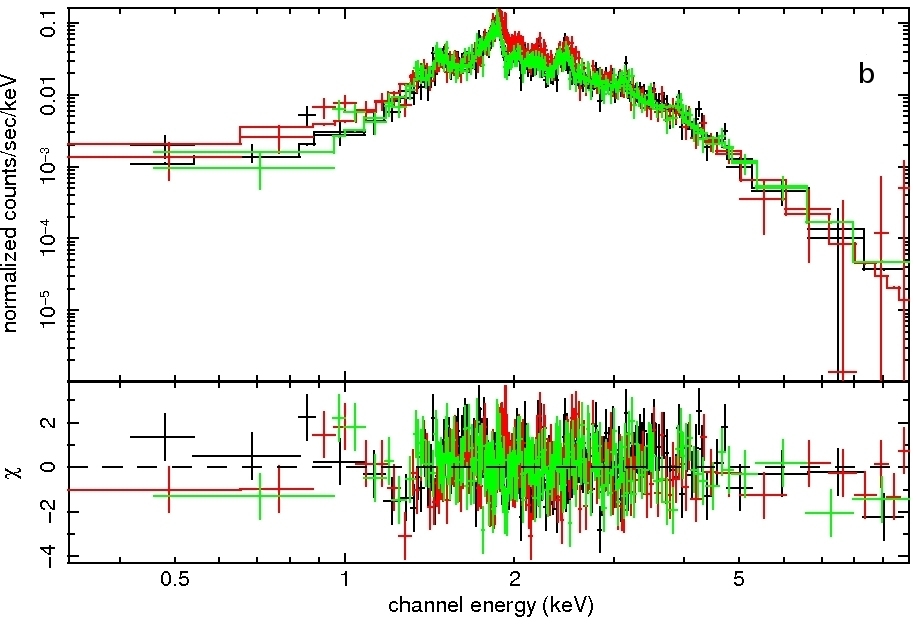}
\includegraphics[width=8cm]{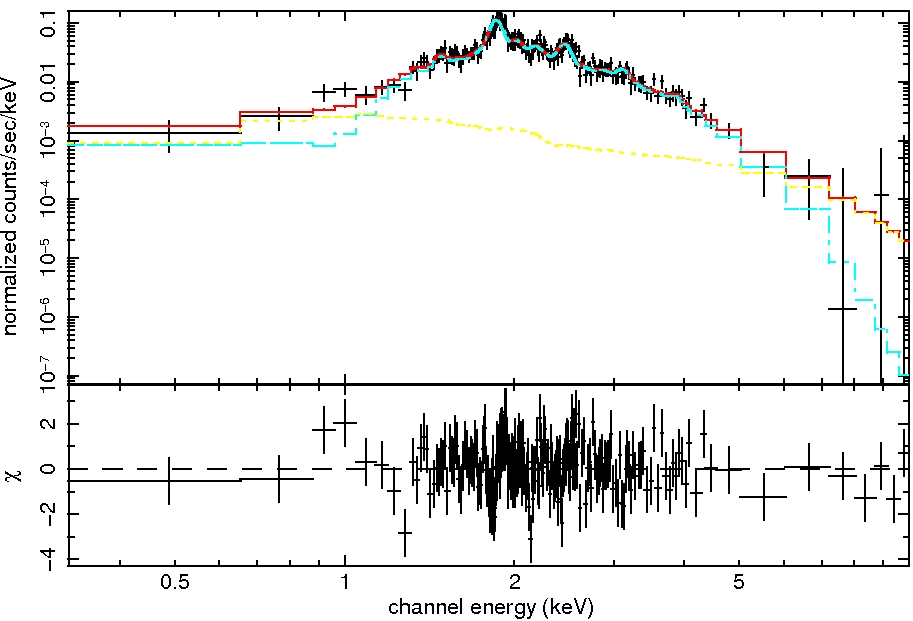}
\includegraphics[width=8cm]{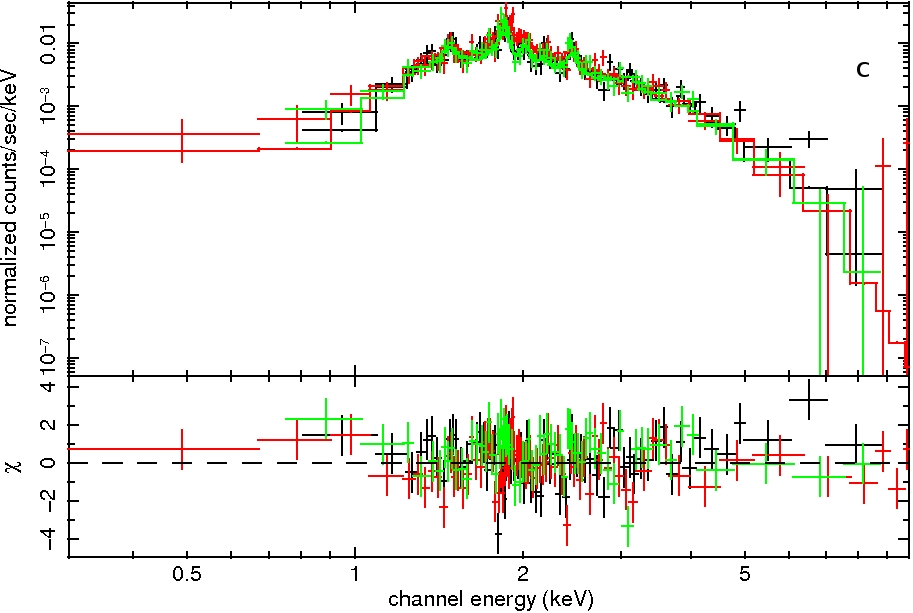}
\caption{Upper left panels: {\it Suzaku} XIS (XIS0:black, XIS1:red
and XIS3:green) spectra have been fitted by an absorbed
  VMEKAL+power-law models in 0.3$-$10 keV energy band simultaneously, a) region 1 and (b) region 2. Upper right panels: The red curve represents the best-fitting of two
  component model (VMEKAL+power-law). The individual components
  are represented by the cyan curve for the VMEKAL model and the
  yellow curve for the power-law model for only XIS1 for region 1 and 2. Bottom panel (c): XIS spectra have been fitted by an absorbed
  VMEKAL model in 0.3$-$10 keV energy band simultaneously for region 3. The lower panels show the residual in units of $\sigma$.}
\end{figure}

\begin{figure}
\centering
  \vspace*{17pt}
\includegraphics[width=8cm]{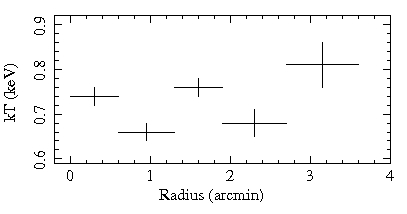}
\includegraphics[width=8cm]{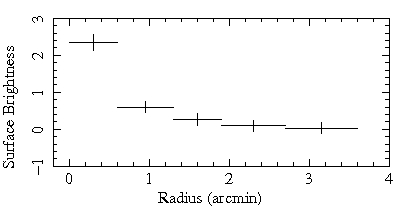}
  \caption{The radial profiles of electron temperature (left panel) and surface brightness (right panel) obtained from XIS0 spectra. Surface brightness
  is in the unit of $\times10^{-12}$ erg $\rm s^{-1}$$\rm
cm^{-2}$$\rm arcmin^{-2}$ in 0.3$-$10 keV energy band.}
\end{figure}

\end{document}